\documentclass[
    ,final            
  ]
  {aipproc}

\layoutstyle{6x9}

\begin{document}

\title{Neutral atom transport from the termination shock to 1 AU}


\author{Maciej Bzowski}{
  address={Space Research Centre PAS, Bartycka 18A, 00-716 Warsaw, Poland}
}

\author{S{\l}awomir Tarnopolski}{
  address={Space Research Centre PAS, Bartycka 18A, 00-716 Warsaw, Poland}
}

\begin{abstract}
Dynamics of H, D, and heavy Energetic Neutral Atoms (ENA) between the
termination shock and 1~AU is discussed in the context of the
forthcoming NASA SMEX mission IBEX. In particular, effects of the
velocity-dependent radiation pressure on atomic trajectories are
considered and ionization losses between TS and 1~AU are studied. It is
shown, among others, that most of the dynamical effects and ionization
losses are induced within a few AU from the Sun, which translates to the
time domain into $\sim 1 - 3$ solar rotations before detection. This
loosens considerably time requirements for tracking the ionization and
radiation pressure history to just prior 3 months. ENA seem excellent
tracers of the processes within the heliospheric interface, with the
transport effects between the termination shock and detector relatively
mild and easy to account for. 
\end{abstract}

\maketitle
\newcommand{\ssr}{Space Sci. Res.} 

\section{Introduction}

This article is a part of reconnaissance for a NASA SMEX mission IBEX
(\cite{mccomas_etal:04a}; McComas et al, this volume). IBEX will be a
spin-stabilized Earth satellite observing Energetic Neutral Atoms (ENA)
in the 0.01 -- 6~keV energy band, with sensors looking perpendicularly
to the spin axis directed at the Sun. The prime target of the mission
will be imaging of the heliospheric interface via neutral atoms and the
goal of an individual observation will be to obtain the kinematic
parameters of the locally registered atom beyond the termination shock
(TS). To that end, understanding the processes affecting the atoms
between their source region and the detector is needed. In particular, a
control on the modifications of the kinematic parameters of the atoms
between the source region and the detector and a grasp on loss processes
underway is necessary.

Since the ENA gas inside TS is collisionless at all energies, then --
from Boltzmann equation -- the transport of ENA from TS to the inner
heliosphere is governed solely by dynamic effects and gain and loss
processes.

The losses include all known ionization processes operating in the
heliosphere: charge exchange, EUV ionization, and electron impact
\cite{rucinski_etal:96a}. For inward-traveling ENA, gains (considered by
IBEX as unwanted background) include charge exchange between CME, CIR
and pickup ions and all kinds of neutral populations present inside TS,
as well as production of ENA in the plasma environment of the spacecraft
(McComas et al., this volume) and will be neglected.

\section{Dynamics of H and D atoms with Doppler effect}

 \begin{figure}
  \includegraphics[height=.3\textheight]{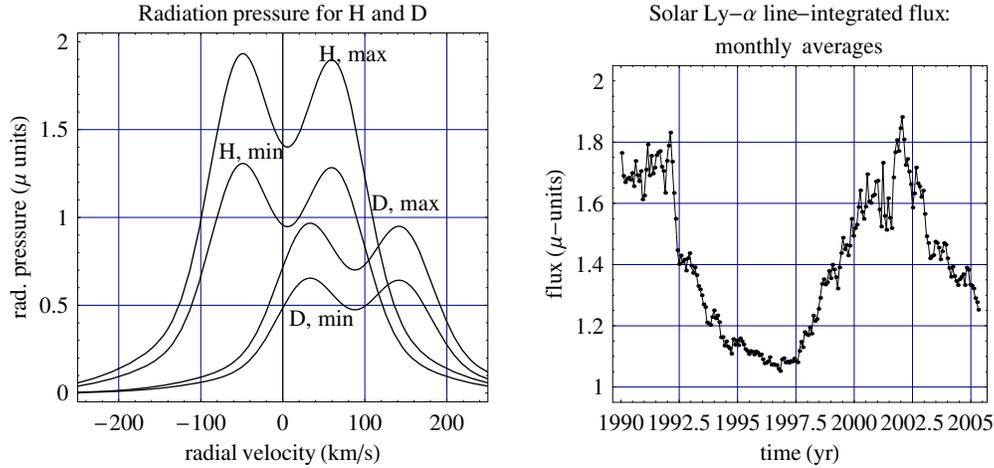}
  \caption{Radiation pressure force acting on H and D atoms as function of atomic
   radial velocity for solar minimum and maximum conditions, expressed in the units
   of solar gravity force (``the $\mu$ units") and
   the net line-integrated solar Lyman-$\alpha$ flux for the past $\sim 1.5$ solar cycle,
   also expressed in the $\mu$ units. }
  \label{fig1}
\end{figure}
From dynamical point of view, heliospheric ENA belong to two categories:
sensitive and insensitive to solar radiation pressure. The sensitive
ones include only hydrogen -- the most abundant heliospheric element --
and its isotope deuterium. All atoms are, of course, sensitive to solar
gravity.

All neutral atoms in the heliosphere obey the equation of motion in
the form:
\begin{equation}
d^2 \vec{r}/{dt^2} = -G\, M \left(1 - \mu\left(v_r, I\left(t\right)\right)\right) \vec{r}/|\vec{r}|^3
\label{equ2}
\end{equation}
where $v_r = \left(d \vec{r}/dt\right) \cdot \vec{r}/|\vec{r}|$ is
radial velocity and $G\,M$ the gravity constant times solar mass. 
$I\left(t\right)$ is the line-integrated solar Lyman-$\alpha$ 
flux. For all atoms heavier than D the radiation pressure factor $\mu 
\equiv 0$ and the equation of motion reduces to the simple Keplerian case, with
the ideal hyperbolic solution.

The expected range of velocities of heliospheric H and D ENA is $\sim 50
- 1000$~km/s. Beacuse of the geometry of the lines of sight, IBEX is
interested only in the inward-traveling atoms, with velocities in the
source region pointing so that at 1~AU they will be at perihelion. Such
atoms are being detected at 0 radial velocity with respect to the Sun,
but cover most of their paths with radial velocities almost equal to
their net velocities. Consequently, if their net velocity is larger than
the spectral range of the solar Lyman-$\alpha$ line, the atoms do not
experience any radiation pressure at all all the way except within a few
AU from the Sun, where before detection they reduce their radial
velocities to 0. Since, however, they are fast, their trajectories will
be weakly modified by solar gravity and hence close to straight lines.
On the other hand, slower ENA (whose velocities remain within the
spectral range of the solar line), will be affected by radiation
pressure, which acts against solar gravity. Consequently, they also will
follow trajectories close to straight lines, except perhaps within a few
AU from the detector.

The behavior of solar Lyman-$\alpha$ line from solar minimum to maximum
has been recently surveyed by \cite{lemaire_etal:02}. Based on these
data it was possible to fit a general formula (Tarnopolski, thesis, in
preparation):
\begin{equation}
\mu\left(v_r, I\right) = \left(a + b\, I\right)\left( \exp\left[-c\,v_r^2\right]
\left(d + f \exp\left[-g \, v_r - h \, v_r^2 \right] + j \exp\left[ k\,v_r - m\, v_r^2\right]\right)
\right)
\label{equ1}
\end{equation}
which describes the radiation pressure of H or D atoms traveling at
radial velocity $v_r$ with respect to the Sun. The formula, shown for H
and D and solar minimum and maximum conditions in the left-hand panel of
Fig.\ref{fig1}, is parametrized by the sole parameter $I$, which is the
line-integrated solar Lyman-$\alpha$ flux, whose variations during the
solar cycle (SOLAR 2000, \cite{tobiska_etal:00c}) are shown in the
right-hand panel of Fig.\ref{fig1}.

Due to isotope effect, the deuterium resonant Lyman-$\alpha$ frequency
is shifted blueward by $\sim 80$ km/s with respect to H. Hence the D
atoms are sensitive to a different part of the solar Lyman-$\alpha$
profile than H atoms and due to the mass difference the radiation
pressure potentially acting on D atoms is additionally reduced two-fold
with respect to H. The two effects combined will yield different
dynamics of the H and D ENA: while radiation pressure for H is
approximately symmetric with respect to 0 radial velocity, this is not
the case for D and the effective radiation pressure acting on the
incoming D atom will be much lower than on a H atom launched from the
termination shock with the same speed (see Fig.\ref{fig1}, left-hand
panel). Consequently, the dynamic behavior of the {\em incoming} D ENA
is expected to be more similar to the behavior of purely-Keplerian heavy
atoms than to that of H atoms.

Results presented in the following part of the paper are based on
numerical solutions of Equ.(\ref{equ2}) with the $\mu$ factor taken from
Equ.(\ref{equ1}), evaluated for two fixed values of $I\left(t\right)$,
appropriate for solar minimum and solar maximum. For heavy atoms the
$\mu$ factor was set to 0. The atoms were tracked from their perihelions
at 1~AU to TS, with velocities at perihelion ranging from 50 to
400~km/s. Along with the trajectory tracking the probability of
ionization for an assumed ionization rate at 1~AU was calculated.

 \begin{figure}
  \includegraphics[height=.3\textheight]{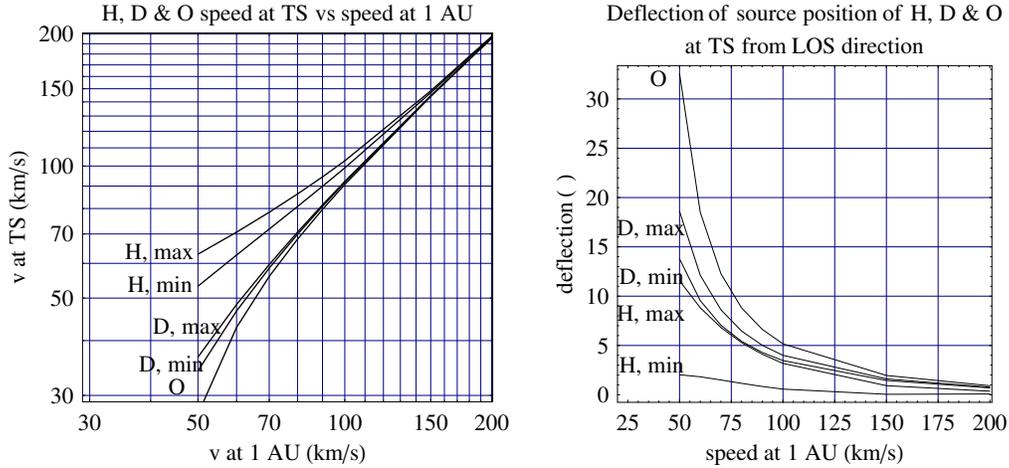}
\caption{Change of speed of H, D, O atoms at solar minimum and maximum
conditions between the termination shock and 1~AU (left-hand panel) and
angular separation between the line of sight (arrival direction at
perihelion) and the source position at the termination shock for H, D, O
atoms at solar minimum and maximum (right-hand panel). The results for O
are valid for all heavy atoms.}
  \label{fig2}
\end{figure}

\section{Results}
\subsection{Trajectories and energy}

All heliospheric ENA except those from the low end of the IBEX energy
band follow almost straight line trajectories, with very little velocity
change between TS and the detector at 1~AU. For detection speeds $v
> \sim 150$~km/s there is practically no velocity change between TS and
1~AU regardless of the species and solar activity phase
(Fig.\ref{fig2}). For $v$ lower than this value, H atoms detected at
velocities $v$ have TS launch velocities $v_{TS,H} > v_{TS,D/O}$, the
launch velocities of the heavier species detected at the same $v$. Also,
usually $v < v_{TS,H}$, while opposite is true for $v_{TS, D/O}$.
Consequently, mean velocity of the H atoms detected at $v$ is higher
than the mean velocities of the heavies and their travel time is
shorter. For example, an O atom detected at 50~km/s needs $\sim 22$~yr
to arrive at 1AU from the crosswind TS, where it was launched at $\sim
26$~km/s, while a H atom detected at the same 50~km/s needs only $\sim
12$~yr to make the trip and it was launched at $\sim 60$~km/s. During solar
minimum, even the slowest H ENA travel almost constant speed, following
straight lines in the whole IBEX energy range. This facilitates
assessment of the travel times $t_{TS}$ from TS to 1~AU for given
detection speed $v$, which can be with good approximation calculated as:
\begin{equation}
t_{TS} = r_{TS}\left(\theta\right)/v = 125.74/\left(1 + 0.253 \cos\, \theta\right)/v
\label{eqn3}
\end{equation}
where $\theta$ is offset angle from the nose of TS and for $t_{TS}$ in yr
$v$ must be in AU/yr. For the lowest-energy atoms ($\sim 100$~km/s)
the difference in travel time from the nose and tail of TS will be
about 5~yr, but for typical H ENA at 400~km/s
it will be only 1~year. The shape of the termination shock was taken
from Grzedzielski et al. (this volume). 

Mapping of the launch position of H and D atoms at TS for a given
detector pointing is practically 1 to 1, with the deflection angle at
the lower end of the energy band within $\sim 10^{\rm o}$
(Fig.\ref{fig2}, right-hand panel). Heavy atoms (insensitive to
radiation pressure) are also weakly deflected except the slowest ones,
which are deflected up to $\sim 30^{\rm o}$ at 50~km/s. Deflection of all 
species drops below $5^{\rm o}$ for detection speed 100~km/s. This is 
favorable for IBEX since it facilitates linking the pointing 
of detectors to the source region at TS. 

Also favorable is the relatively short range of the combined solar
gravity and solar radiation pressure acting on H ENA. The effective
``collision time'' of H/D ENA with the Sun before detection is only
$\sim 0.25\,{\rm yr} = 3$~solar rotations, so the highly fluctuating
solar Lyman-$\alpha$ output will need to be tracked back no longer than
this interval. 

\subsection{Ionization}
 \begin{figure}
  \includegraphics[height=.3\textheight]{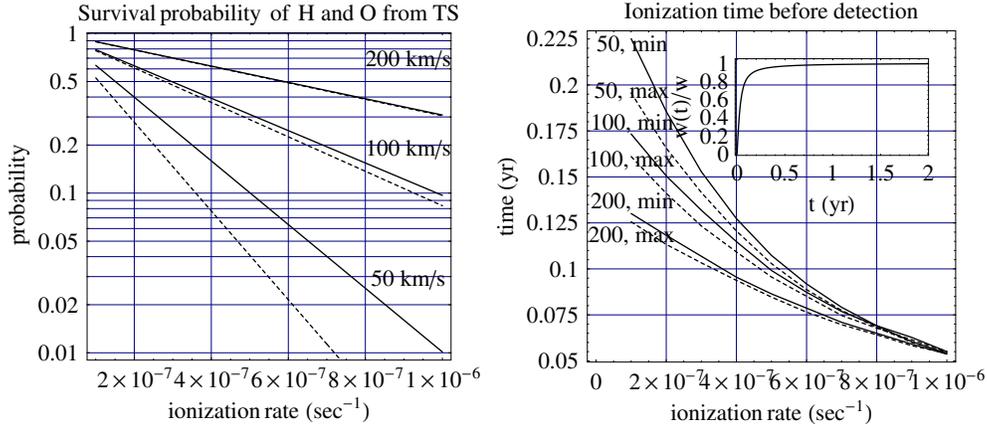}
\caption{Left-hand panel: Survival probability of H (solid lines) and O
ENA (broken lines) from TS to 1~AU as function of the ionization rate at
1~AU for various detection speeds. The solid lines in the left-hand
panel correspond to H ENA, the broken lines to heavy atoms (like O); for
200 km/s the two lines practically coincide. Right-hand panel: time of
effective ionization defined in Equ.(\ref{eqn4}) for H at solar minimum
and maximum conditions as function of the ionization rate at 1~AU
(right-hand panel) for various detection speeds, indicated in the figure,
and for solar minimum (solid) and maximum conditions(broken). 
The inset in the right-hand panel shows an example
of the course of relative survival probability $w_{rel}\left(t\right)$
between TS and $r$.}
  \label{fig3}
\end{figure}
The survival probability of ENA against ionization between TS and 1~AU
is a function of the launch and detection speeds and of the ionization
rate (assumed here to vary with heliocentric distance as $1/r^2$). For
the H ENA expected to be observed by IBEX it is quite large in the whole
energy band except at the lowest end: for a H atom detected at 200~km/s
even for the highest expected ionization rate $10^{-6}$~s$^{-1}$ 1~AU
\cite{rucinski_etal:96a} the probability is as much as 40\%, and for the
lowest expected rate equal to $10^{-7}$~s$^{-1}$ it is almost 1. For
faster H, D, and heavy atoms the survival probability is correspondingly
larger, depending on the ionization rate. On the other hand, the
ionization losses will increase quickly with the decrease of detection
speed, as shown in the left-hand panel of Fig.\ref{fig3}. In the range
of detection speeds shown, one expects some ``radiation pressure bonus''
in the survival probability for the H atoms (and to some extent for D)
compared with the heavy atoms because of the larger mean speed of H
atoms detected at a given $v$ in comparison with the mean speed of heavy
atoms. This effect is significant only at the lowest end of the
detection interval. For H, some variations of the survival probability
with the phase of the solar cycle are expected, but they are not big and
probably will be masked by fluctuations of the actual ionization rate.

Most of the ionization losses occur shortly before detection. Denoting
by $w_{\rm 1AU}$ the ionization probability of an atom underway from TS
to 1~AU and by $w\left(t\right)$ the probability of ionization at time
$t$ before detection, we can define the relative survival probability
$w_{rel}\left(t\right)$ and the ionization time $\tau_{ion}$ as:
\begin{equation}
w_{rel}\left(t\right) = 1 - w\left(t\right)/w_{\rm 1AU};\, \,  w_{rel}\left(\tau_{ion}\right) \equiv 0.8
\label{eqn4}
\end{equation}
and determine the ionization times for different species and detection
speeds. Results are shown in Fig.\ref{fig3} (right-hand panel). The
ionization losses suffered by an atom underway from TS to 1 AU are
relatively small almost all the way down, and increase exponentially
only within few months before detection. The exact value of $\tau_{ion}$
is a function of detection speed and ionization rate, but generally one
can say that the 80\% ionization losses of all atoms occur within 0.1 --
0.25~yr ($\sim 1 - 3$ solar rotations) before detection. The ionization
times differ most for the lowest expected ionization rates and for the
lowest energies at detection; for typical H ENA expected from the inner
heliosheath the ionization time is expected at the level of one solar
rotation period. This is favorable for IBEX operations because it means
that interpretation of a signal requires monitoring of the local
ionization rate only $\sim 1$~month backward in time. 

\section{Conclusions}

The aspects of ENA transport from the termination shock to 1~AU in the
contexts of the planned IBEX mission can be summarized as follows.

Both heavy ENA and H and D ENA are excellent tracers of atomic
parameters at and beyond TS. Velocity change between the termination
shock and 1~AU becomes noticeable only below 150~km/s and deflection of
the trajectories with respect to the launch site at TS beyond angular
resolution of IBEX instruments for H ENA occurs only during solar
maximum (which will be beyond IBEX nominal mission time) at the lowest
limit of the energy band. Heavy atoms are a little stronger deflected,
but the deflections exceed IBEX resolution only for detection speeds lower
than $\sim 70$~km/s. 

Ionization losses between TS and 1~AU should not be a big problem for 
H ENA expected from the heliosheath. The survival probability drops down 
below 10\% only for H ENA traveling at 1~AU slower than 100~km/s. Most
of the ionization losses will occur within a month or so before detection
and hence should be relatively easy to take into account.


\begin{theacknowledgments}
This research was supported by the Polish MSRiT Grant 1P03D 009 27.
\end{theacknowledgments}


\bibliographystyle{aipproc}   

\bibliography{iplbib}

\begin{thebibliography}{4}
\expandafter\ifx\csname natexlab\endcsname\relax\def\natexlab#1{#1}\fi
\providecommand{\enquote}[1]{``#1''}
\expandafter\ifx\csname url\endcsname\relax
  \def\url#1{\texttt{#1}}\fi
\expandafter\ifx\csname urlprefix\endcsname\relax\def\urlprefix{URL }\fi
\providecommand{\eprint}[2][]{\url{#2}}

\bibitem[{McComas} et~al.(2004)]{mccomas_etal:04a}
D.~{McComas}, F.~{Allegrini}, P.~{Bochsler}, M.~{Bzowski}, M.~{Collier},
  H.~{Fahr}, H.~{Fichtner}, P.~{Frisch}, H.~{Funsten}, S.~{Fuselier},
  G.~{Gloeckler}, M.~{Gruntman}, V.~{Izmodenov}, P.~{Knappenberger}, M.~{Lee},
  S.~{Livi}, D.~{Mitchell}, E.~{M{\" o}bius}, T.~{Moore}, D.~{Reisenfeld},
  E.~{Roelof}, N.~{Schwadron}, M.~{Wieser}, M.~{Witte}, P.~{Wurz}, and
  G.~{Zank}, \enquote{{The Interstellar Boundary Explorer (IBEX)},} in
  \emph{AIP Conf. Proc. 719: Physics of the Outer Heliosphere}, 2004, pp.
  162--181.

\bibitem[Ruci{\'n}ski et~al.(1996)]{rucinski_etal:96a}
D.~Ruci{\'n}ski, A.~C. Cummings, G.~Gloeckler, A.~J. Lazarus, E.~M{\"o}bius,
  and M.~Witte, \emph{\ssr} \textbf{78}, 73--84 (1996).

\bibitem[Lemaire et~al.(2002)]{lemaire_etal:02}
P.~L. Lemaire, C.~Emerich, J.~C. Vial, W.~Curdt, U.~Sch{\"u}le, and K.~Wilhelm,
  \enquote{Variation of the full Sun hydrogen Lyman $\alpha$ and $\beta$
  profiles with the activity cycle,} in \emph{ESA SP-508: From Solar Min to
  Max: Half a Solar Cycle with SOHO}, 2002, pp. 219--222.

\bibitem[{Tobiska} et~al.(2000)]{tobiska_etal:00c}
W.~K. {Tobiska}, T.~{Woods}, F.~{Eparvier}, R.~{Viereck}, L.~{Floyd},
  D.~{Bouwer}, G.~{Rottman}, and O.~R. {White}, \emph{J. Atm. Terr. Phys.}
  \textbf{62}, 1233--1250 (2000).

\end{thebibliography}

\end{document}